\documentclass[aps,prx,citeautoscript,showpacs,showkeys,groupedaddress,superscriptaddress,twocolumn]{revtex4-2}

\usepackage[utf8]{inputenc}
\usepackage{color}
\usepackage{xcolor}
\usepackage{bbm} 
\usepackage{amsfonts,amsmath,amssymb,stmaryrd}
\usepackage{graphicx}
\usepackage{subcaption}  
\usepackage{hyperref}
\usepackage{epsfig}
\usepackage{mathrsfs}
\usepackage{verbatim}
\usepackage{centernot}
\usepackage{svg}
\usepackage{hyperref}
\usepackage{epsfig}
\usepackage{mathrsfs}
\usepackage{verbatim}
\usepackage{centernot}
\usepackage{siunitx}
\usepackage{braket}
\usepackage{wasysym}
\usepackage{interval}
\intervalconfig{soft open fences}
\usepackage{mathtools}
\usepackage{amsthm}
\usepackage{float}
\usepackage{mhchem} 
\usepackage[normalem]{ulem}
\usepackage{dirtytalk}

\newcommand{\hc}{\text{h.c.}}

\newcommand{\nn}[1]{\langle #1 \rangle}
\newcommand{\nnn}[1]{\langle \nn{#1} \rangle}
\newcommand{\up}{\uparrow}
\newcommand{\down}{\downarrow}

\usepackage{array}

\usepackage{bm}	
\renewcommand{\vec}[1]{\bm{#1}}

\begin{document}

\title{
Parton Mean-Field Theory of a Rydberg Quantum Spin Liquid induced by Density-Dependent Peierls Phases
}

\author{Benno Bock}
\affiliation{Department of Physics and Research Center OPTIMAS, RPTU-University of Kaiserslautern-Landau, D-67663 Kaiserslautern, Germany}
\author{Simon Ohler}
\affiliation{Department of Physics and Research Center OPTIMAS, RPTU-University of Kaiserslautern-Landau, D-67663 Kaiserslautern, Germany}
\author{Michael Fleischhauer}
\affiliation{Department of Physics and Research Center OPTIMAS, RPTU-University of Kaiserslautern-Landau, D-67663 Kaiserslautern, Germany}

\begin{abstract}
We derive a parton mean-field Hamiltonian for Rydberg excitations on a honeycomb lattice with nearest and density-dependent, complex next-nearest neighbor hopping. Numerical results obtained from exact diagonalization of small systems have given indications for a ground state that is a chiral spin liquid (CSL)
[Phys.Rev.Res. \textbf{5}, 013157 (2023)].
Here we provide further evidence for this.
Calculating the ground-state wavefunction self-consistently, we show that the mean-field Hamiltonian fulfills the requirements for a CSL ground state, resulting from a projected symmetry group classification and verify the expected twofold topological degeneracy on a torus. Furthermore we find very good overlap with the ground-state wavefunctions obtained by exact diagonalization of the original Hamiltonian.
\end{abstract}


\date{\today}
\maketitle

\section{Introduction}\label{sect:introduction}

Quantum spin liquids (QSLs) are fascinating states of matter that are primarily characterized by a lack of any magnetic order in the ground state \cite{anderson1973resonating,Anderson-Science_1987,knolle2019field}. Quantum fluctuations prevent the system to be ordered even at zero temperature.
In recent years, other properties of QSLs such as massive many-body entanglement have moved into focus \cite{savary2017quantum}. While theoretically the existence of quantum spin liquids is certain, such states being the solution to a growing number of exactly solvable models \cite{kitaev2006anyons,fu2019exact,ben2016exactly}, the quest for finding real materials that exhibit such exotic phases is still a formidable challenge \cite{mila2000quantum,lee2008end,savary2017quantum,knolle2019field,broholm2020quantum,gohlke2018quantum,liu2022gapless,coldea2001experimental,shen2016evidence}.

Recently, quantum simulators based on Rydberg atoms have emerged as platforms to engineer Hamiltonians that can host spin liquid ground-states \cite{semeghini2021probing,giudici2022dynamical,chen2024proposal}. Such platforms possess the advantage that the microscopic interactions between the individual spins can be tightly controlled and tuned to mimic, e.g., quantum dimer models, where Rydberg blockade enforces the dimer constraint \cite{semeghini2021probing}.

In \cite{ohler2023quantum} some of the authors of this work presented a model of interacting Rydberg atoms on a honeycomb lattice that included a density-dependent Peierls phase responsible for chiral motion of the Rydberg excitations. Numerical calculations and a mapping to fermions produced evidence that frustration in this model leads to a chiral spin liquid (CSL) state. Crucially, however, no ground-state degeneracy was found on the torus, likely due to finite-size limitations. In a subsequent publication \cite{tarabunga2023classification}, the observation of a QSL-phase was further solidified using the projective symmetry group (PSG) classification of CSLs \cite{bieri2016projective}, based on a parton mean-field theory of spin liquids \cite{Wen2002,wen2002quantumPLA}. The authors considered different ansatz Hamiltonians derived from general symmetry considerations, 
 selected the best fitting model, and determined its parameters by comparison with exact diagonalization (ED) results for small systems.
However, the physical origin of the ansatz Hamiltonian remained unclear and the corresponding parameter were ad-hoc fits.

In this work, we derive the parton mean-field Hamiltonian from microscopic properties and compute the ground state self-consistently. The self-consistent solution verifies the specific form of the Hamiltonian suggested in \cite{tarabunga2023classification} and shows remarkable agreement with ED-simulations. In addition it yields a twofold topological degeneracy expected for CSLs but not observed in ED. 

\section{Model Hamiltonian}\label{sec:model-hamiltonian}

In this work we consider the model originally proposed in \cite{ohler2023quantum}, derived from microscopic dipole-dipole interactions of Rydberg atoms.

\begin{figure}[H]
	\centering
	\includegraphics[width=\columnwidth]{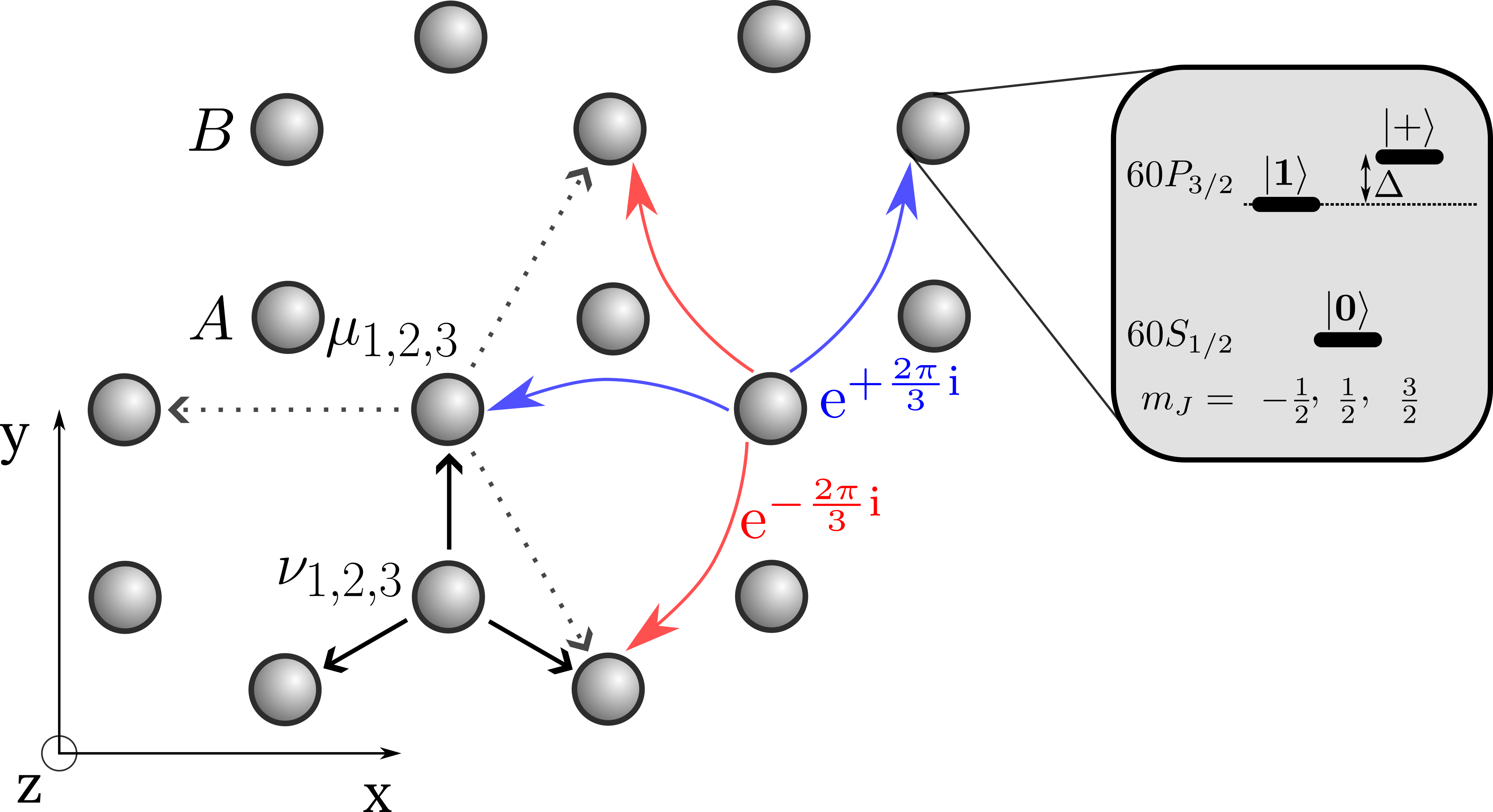}
	\caption{Honeycomb lattice with a two-site unit cell (A and B) of trapped atoms excited to Rydberg states $\ket{1}$ and $\ket{0}$, forming spin-$1/2$ systems. Spin-orbit coupling induced by an external magnetic field leads to nonlinear, complex second-order hopping processes to the next-nearest neighbor (NNN, curved arrows) in addition to direct nearest neighbor (NN, black arrows) hopping. The relevant level structure of a single atom is shown in the inset. The NNN hopping is facilitated by virtual transitions from $\vert 0\rangle$ to the off-resonant state $\vert +\rangle$ detuned by $\Delta$. The black arrows and dashed arrows correspond to the lattice vectors $\vec{\nu}, \vec{\mu}$.
	}
	\label{fig:RydbergHaldane_allHopping}
\end{figure}

As can be seen in Fig.~\ref{fig:RydbergHaldane_allHopping}, the atoms are placed on the nodes of a honeycomb lattice at nearest neighbor (NN) distance $r$, where every atom is characterized by three energy levels: the qubit (or spin-$1/2$) states $\ket{0}$ and $\ket{1}$ as well as the auxiliary state $\ket{+}$, which represent the Zeeman-split sublevels of a Rydberg $S$ and $P$ manifold. The interactions between the Rydberg atoms are then composed of several components. Firstly, the simple dipole-dipole exchange process leads to a NN hopping of the Rydberg $P$ excitation (state $\ket{1}$). However, due to the auxiliary state $\ket{+}$ one also obtains non-resonant processes, which lead to a second hopping process that is associated with a Peierls phase and depends on the occupation of a third atom. This Hamiltonian has been studied in previous publications \cite{lienhard2020realization,weber2018topologically,ohler2023quantum,ohler2022self}, which is why we refer to those for a detailed derivation. Adiabatically eliminating the $\ket{+}$ state and setting $J=d^2/(8\pi\epsilon_0r^3)$, with $d$ being the dipole matrix element between $\ket{0}$ and $\ket{1}$, as the unit of energy the Hamiltonian reads
\begin{align}\label{eqn:Rydberg_BHH_fullH}
\hat{H}=
&
-J\sum_{\langle i,j\rangle}\hat{b}_{j}^{\dagger}\hat{b}_{i}
-2gJ\sum_{\langle\langle i,j\rangle\rangle}\hat{b}_{j}^{\dagger}\hat{b}_{i}\mathrm{e}^{\pm\frac{2\pi\mathrm{i}}{3}}(1-\hat{n}_{ij})
\nonumber\\
&+2gJ\sum_{\langle i,j\rangle}\hat{n}_{i}\hat{n}_{j} + \hc
\end{align}
$\hat{b}^\dagger_i$ and $\hat{b}_i$ create or destroy a hard-core boson on lattice site $i$, respectively, $\hat{n}_i=\hat{b}^\dagger_i\hat{b}_i$ is the number operator. We label the two triangular sublattices of the honeycomb lattice $A$ and $B$ respectively.
We define $\nn{ij}$ to signify summation over NNs with $i \in A$ and $\nnn{ij}$ to mean summation over next-nearest neighbors (NNNs) where $i \to j$ is in counterclockwise  direction.
\\

In Fig.~\ref{fig:RydbergHaldane_allHopping} the different hopping terms are illustrated. Black solid arrows correspond to NN hopping, while the red and blue arrows signal the NNN terms with the complex phase encoded in the color. 
The respective intermediate site, which controls the hopping, is indicated in the bend of the arrow. 
The dashed black arrows denote the lattice vectors used later.
While the NN hopping is of constant strength, the NNN hopping as well as the NN density-density interaction terms scale with the parameter $g= c\, J/\Delta$, which is determined by the detuning of the auxiliary $\ket{+}$ state. Since $\Delta$ has to be large to justify the adiabatic elimination, the value of $J/\Delta$ has to remain small, however $g\sim 2$ is possible, due to favorable Franck-Condon factors contained in c.

The mean-field approximation to Hamiltonian 
\eqref{eqn:Rydberg_BHH_fullH}, obtained by replacing the number operators by their expectation value at half filling,
\begin{equation}
\hat{H}=
-J\sum_{\langle i,j\rangle}\hat{b}_{j}^{\dagger}\hat{b}_{i}
-gJ\sum_{\langle\langle i,j\rangle\rangle}\hat{b}_{j}^{\dagger}\hat{b}_{i}\mathrm{e}^{\pm\frac{2\pi\mathrm{i}}{3}} + \hc
\end{equation}
contains two competing terms, which lead to frustration at $g\approx 0.5$. This can be seen most easily in an equivalent spin representation 
$\hat b_i \sim \hat S_i^-$, and $\hat b_i^\dagger \sim \hat S_i^+$: The NN hopping aims to align the spins parallel in the $xy$-plane, while the $g$-dependent NNN hopping term prefers a $120^\circ$ orientation on each sublattice. The competition between the two causes a phase transition at $g\approx 0.5$. Close to this point quantum fluctuations of the term $\hat n_{ij}$ become relevant and induce an intermediate phase that is characterized by the absence of spin order as well as the breaking of chiral symmetry, manifested by a Chern number of $C=1$, found by ED simulations in \cite{ohler2023quantum}, and is thus a candidate for a CSL state.

However a definite identification of unambiguous spin liquid characteristics such as the topological entanglement entropy and in particular the ground-state degeneracy were not obtained in the numerics. Chiral spin liquids are expected to be described in the low-energy regime by a Chern-Simons theory \cite{wen1990topological,bieri2016projective}, which implies a ground-state degeneracy depending on the genus $g$ of the compactified parameter space. In the simplest case of an Abelian theory the degeneracy is $2^g$, i.e. $2$ on a torus.
Using projective symmetry group (PSG) arguments and ansatz Hamiltonians, the authors of \cite{tarabunga2023classification} determined the putative spin liquid phase to indeed be a CSL state.

In this publication, we expand on the previous results and construct a parton Hamiltonian not from symmetry considerations but directly from the microscopic physics of the hard-core bosonic Rydberg excitations. We obtain an analytical ground-state wavefunction and compute physical observables confirming the accuracy of our approach.

\section{Parton construction and mean-field approach}
Since QSLs are characterized by vanishing magnetic correlations $\langle\hat{\vec{S}}_i \rangle= 0$, it is not possible to do a straight-forward mean-field treatment on the spin-operator basis.
However, a way to construct wavefunctions of 2D spin liquids is the projective construction by Wen \cite{Wen2002} first introduced in the context of high-$T_c$ superconductors by Baskaran, Zou and Anderson \cite{BASKARAN1987973}. 
One introduces fermionic parton operators $f_{i,\alpha}$ called spinons with $\alpha \in \{\uparrow, \downarrow\}$, which carry spin $\frac{1}{2}$ and replace spin operators by
\begin{equation}
\label{eq:fermionic-spin-operators}
  \hat{\vec{S}} =\frac{1}{2} f^\dagger_\alpha\vec{\sigma}_{\alpha\beta} f_\beta ,  
\end{equation}
where $\vec{\sigma}$ is the vector of Pauli matrices (e.g. ${\hat{S}_x =\frac{1}{2}( f^\dagger_{\up}f_{\down} + f^\dagger_{\down}f_{\up})}$).
It is now possible to do a mean-field decoupling of the Hamiltonian where all expectation values corresponding to magnetic order can be set to zero \cite{PhysRevB.90.174417}, i.e.
\begin{equation}
    \langle \hat{\vec{S}_i} \rangle= \frac{1}{2}
    \langle 
    f^\dagger_{i,\alpha}\vec{\sigma}_{\alpha\beta} f_{i,\beta} 
    \rangle
    =0.\label{eq:av-S}
\end{equation}
The fermion-parton replacement enlarges the Hilbert space to contain unphysical states where a site is occupied by zero or two fermions. This fact is dealt with in the end by performing a Gutzwiller-Projection on the resulting mean-field ground state where these unphysical states are projected out, i.e. $\hat{P}_G = \Pi_i \hat{n}_i(2- \hat{n}_i)$.

For our mean-field model we first replace the bosonic operators in eq. \eqref{eqn:Rydberg_BHH_fullH} according to eq.\eqref{eq:fermionic-spin-operators}, i.e. a bosonic creation operator is replaced by the product of a creation operator of a spin-up and an annihilation operator of a spin-down fermion. The resulting Hamiltonian is of sixth order in the parton operators.

We then perform a mean-field decoupling of the resulting parton-Hamiltonian.
We simplify the problem by setting the expectation value for pair creation and annihilation to zero:
\begin{equation}
    \eta_{ij}\epsilon_{\alpha\beta} = \langle f_{i,\alpha}f_{j,\beta}\rangle \overset{!}{=}0
    \label{eq:pairing-terms}
\end{equation}
This approximation is justified since we want to obtain the ground-state at half filling, which corresponds to a double half filling of the fermion-partons, and the operation in eq. \eqref{eq:pairing-terms} moves a state out of this particle-number conserved subspace.

This leaves only the expectation value for fermion-hopping
\begin{equation}
\chi_{ij, \alpha}= \langle f^\dagger_{i,\alpha}f_{j,\alpha}\rangle
\label{eq:fermion-hopping}
\end{equation}
with no implicit summation over $\alpha$, which is restricted to NN and NNN hopping and which we call $\chi_{1, \alpha}$ and $\chi_{2,\alpha}$ respectively.
Note that we allow for the fermion hopping of the two species $\up$ and $\down$ to be independent of each other, in contrast to the procedure in \cite{Wen2002}.
 
As usual in Wen's method, one expresses the mean-field Hamiltonian in terms of the spinor operator ${\Psi = \left(f_\uparrow, f^\dagger_{\downarrow}\right)^T}$. 
Restricting this kind of mean-field Hamiltonian to NN and NNN interaction leads to
\begin{equation}
    H_{MF}(\chi_{1,2}) = \sum_{\nn{ij}}\Psi^\dagger_i V \Psi_j + \sum_{\nnn{ij}}\Psi^\dagger_i W\Psi_j + \hc
    \label{eq:MF-Hamiltonian-spinor-form}
\end{equation}
where $V$ and $W$ are $2\times 2$-matrices.  
In our case $V$ and $W$ contain no off-diagonal terms due to suppression of spin-pairing terms (see eq. \eqref{eq:pairing-terms}). Thus the MF-Hamiltonian becomes diagonal in the $\up$- and $\down$-fermions.
We furthermore perform a particle-hole-transformation on the $\down$-fermions, which makes it easier to project the parton-state back to the physical subspace: 
\begin{align}
    f_{i,\up}\to \widetilde{f}_{i,\up}=f_{i,\up}  \\
    f_{i,\down} \to \widetilde{f}^\dagger_{i,\down}=f_{i,\down}.
\end{align}
We can then express the mean-field Hamiltonian (dropping constant terms and setting $J=1$) as
\begin{equation}
    \begin{split}
        &H_{MF}(\{\widetilde{\chi}_{1, \alpha}, \widetilde{\chi}_{2,\alpha}\}) = \\
        &\sum_{\alpha=\up,\down}  \left[\sum_{\nn{ij}} v_\alpha \widetilde{f}_{i, \alpha}^\dagger\widetilde{f}_{j, \alpha} \right. 
         + \left. \sum_{\nnn{ij}} w_\alpha\widetilde{f}_{i, \alpha}^\dagger\widetilde{f}_{j, \alpha} \right] + \hc .
    \end{split}
    \label{eq:final-mf-hamiltonian}
\end{equation}
With 
\begin{align}
    &{{v}_{\alpha} = -\widetilde{\chi}_{1,  (-\alpha)} - g\widetilde{\chi}^{\dagger}_{1, \alpha}\left(4\mathrm{Re}(e^{-i\frac{2\pi}{3}}\widetilde{\chi}_{2,(-\alpha)} ) + 1\right)},
    \\
    &{{w}_{\alpha} = -ge^{-i\frac{2\pi}{3}}\left(\widetilde{\chi}_{2,(-\alpha)} + |\widetilde{\chi}_{1, (-\alpha)}|^2\right)}
\end{align}
where $(-\alpha)$ denotes the opposite spin-index.

\begin{figure}[H]
    \centering
    \includegraphics[width=.8\columnwidth]{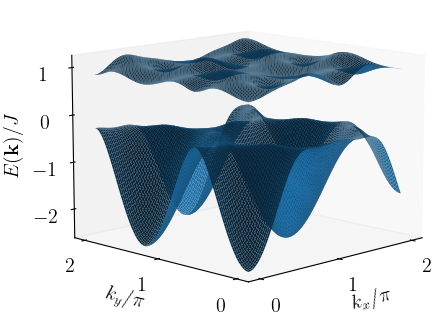}
    
    \caption{Spectrum of $\up$-fermions for self consistent mean-field values at $g=0.7$}
    \label{fig:spectrum}
\end{figure}

Note that although $H_{MF}$ is quadratic in the fermionic operators, it is still a nonlinear function of the correlations $\widetilde{\chi}_{1/2, \alpha}$ via eq.~\eqref{eq:fermion-hopping}.
\begin{figure}[H]
    \centering
    \includegraphics[width=0.9\columnwidth]{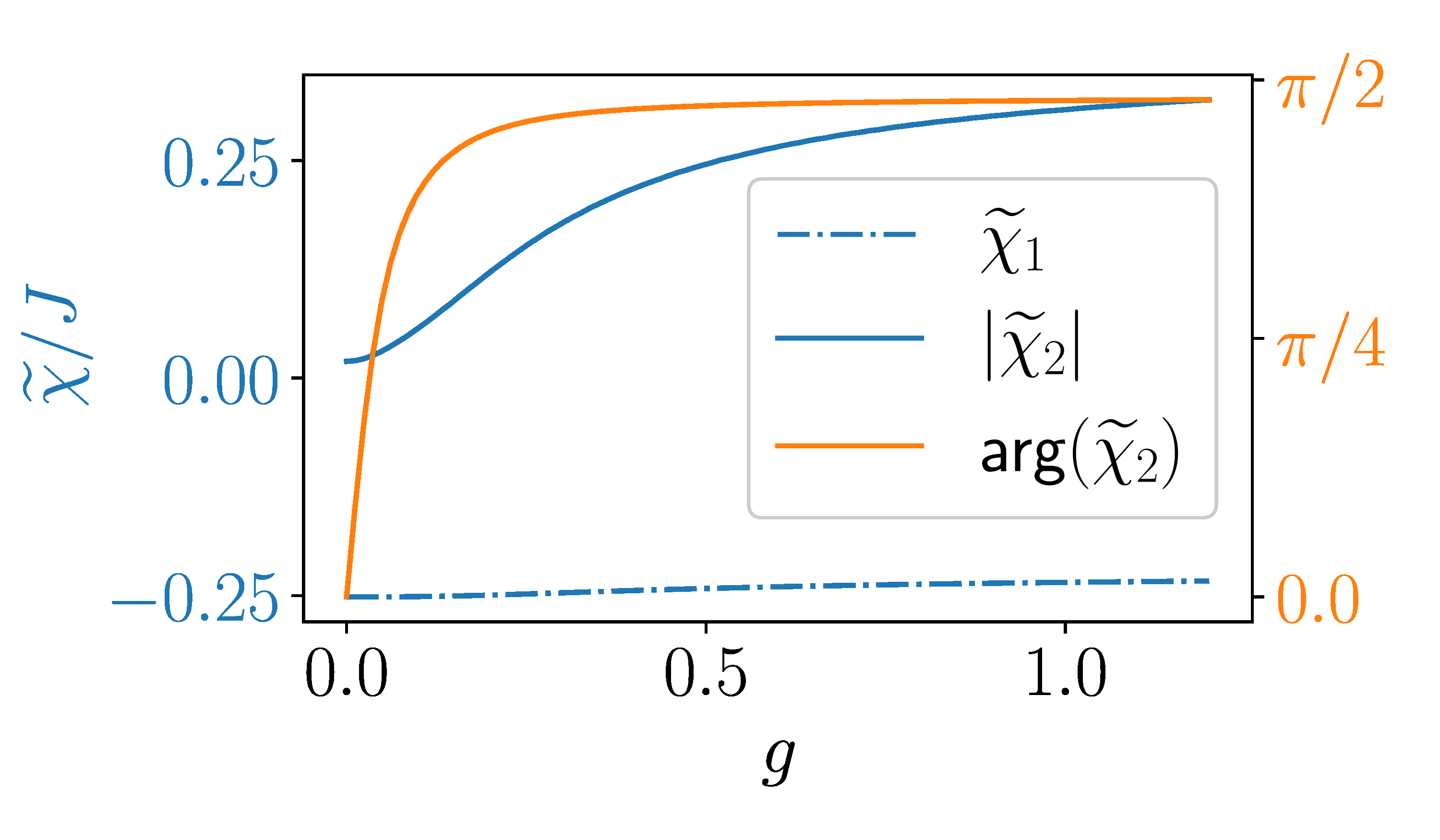}
    \caption{Self consistent solutions of (real) NN- ($\chi_1$) and (complex) NNN- ($\chi_2$) correlations as function of $g$.}    
    \label{fig:Hopping-elements}
\end{figure}
Because of this we need to find the ``true" ground state by self-consistently determining $\widetilde{\chi}_{1/2,\alpha}$ as the limiting values of the following iteration scheme:
\begin{align}
    \widetilde{\chi}^{(n+1)}_{1} &=\frac{2}{3N}\bra{\psi^{(n)}_{MF}}\sum_{\nn{ij}}\widetilde{f}_{i}^\dagger\widetilde{f}_{j}\ket{\psi^{(n)}_{MF}}, \label{eq:discrete-iteration-scheme-1}\\
    \widetilde{\chi}^{(n+1)}_{2} &=\frac{1}{3N}\bra{\psi^{(n)}_{MF}}\sum_{\nnn{ij}}\widetilde{f}_{i}^\dagger\widetilde{f}_{j}\ket{\psi^{(n)}_{MF}} \label{eq:discrete-iteration-scheme-2}.
\end{align}
where $N$ denotes the number of lattice sites. In the limit of $N\to \infty$ we can express each of these expectation values analytically. 
To this end, we note that $H_{MF}$ is diagonal in the two spinon species, so we can find the ground state for each spinon separately.
Each spectrum has two bands (depicted in Fig.~\ref{fig:spectrum}) since the honeycomb lattice features a two-site unit cell.
The Hamiltonian in each spin-subspace can be diagonalized in momentum space by a unitary transformation that can be represented by a $2\times2$-matrix
\begin{equation}
    U_k = \begin{pmatrix}
        a(\vec{k}) & c(\vec{k})\\
        b(\vec{k}) & d(\vec{k})
    \end{pmatrix}
\end{equation}
so that $U^\dagger_kH_kU_k=H_{diag}$, where we drop the spin-index $\alpha$ for simplicity. Since $U_k$ is a function of the mean-field Hamiltonian eq. \eqref{eq:final-mf-hamiltonian} in general $ U_k = U_k (\{\widetilde{\chi}_{1,\alpha},  \widetilde{\chi}_{2,\alpha}\})$.
The ground-state $\ket{\psi_{MF}}$ for half filling of the original bosons is then obtained by filling the lower bands for both $\up$- and $\down$-fermions.
Thus, we can re-express eqs. \eqref{eq:discrete-iteration-scheme-1} and \eqref{eq:discrete-iteration-scheme-2} as

\begin{align}
    \widetilde{\chi}^{(n + 1)}_{1} &=\frac{1}{3A}\int\limits_{\vec k\in1.BZ}\mathrm{d}^2 k\,\varphi_1(\vec{k})({a^{(n)}} (\vec{k}))^*c^{(n)}(\vec{k})\label{eq:iteration-scheme-1}\\
    \widetilde{\chi}^{(n + 1)}_{2} &=\frac{2}{3A}\int\limits_{\vec k\in1.BZ}\mathrm{d}^2 k \,
    \left\{\begin{aligned}
        \varphi_2(\vec{k})|a^{(n)}(\vec{k})|^2 \\
        +\; \varphi^*_2(\vec{k})|c^{(n)}(\vec{k})|^2
    \end{aligned}\right\}
    \label{eq:iteration-scheme-2}
\end{align}
where $A$ denotes the area of the first Brillouin zone (1.BZ) and 
\begin{align}
    \varphi_1 (\vec k) = \sum_{j=1}^3e^{i\vec k \cdot \vec \nu_j},\qquad
    \varphi_2 (\vec k) = \sum_{j=1}^3e^{i\vec k \cdot \vec \mu_j}
\end{align}
where $\vec \nu_j$ are the vectors pointing at NNs from an atom on sublattice $B$  and $\vec \mu_j$ are vectors pointing at NNNs on clockwise direction starting from an atom on sublattice $A$ (see Fig.~\ref{fig:RydbergHaldane_allHopping}).
\
According to \cite{Wen2002}, $\chi_{ij, \up} = \chi_{ij, \down}$. 
In the particle-hole transformed picture this translates to 
\begin{equation}
 \widetilde{\chi}_{ij, \up} = -\widetilde{\chi}^{\dagger}_{ij,\down}  . 
 \label{eq:ph-transformed-hopping}
\end{equation}

To self-consistently determine the correlations $\chi_{1/2, \alpha}$, for each value of $g$, we draw the starting values $\widetilde{\chi}^{(n=0)}_{1,\up}$ and $\widetilde{\chi}^{(n=0)}_{2,\up}$ randomly from the interval $(0,1)$ and set $\widetilde{\chi}^{(n=0)}_{1/2,\down}$ according to eq. \eqref{eq:ph-transformed-hopping}.
We then determine ${\widetilde{\chi}}_{1/2,\up}$ and $ \widetilde{\chi}_{1/2, \down}$ independently of each other according to eqs. \eqref{eq:iteration-scheme-1} and \eqref{eq:iteration-scheme-2}.
Remarkably, the iteration series converges to a point where ${\widetilde{\chi}}_{1, \up} = \widetilde{\chi}_{1, \down}\coloneq \widetilde{\chi}_{1}$ and ${\widetilde{\chi}}_{2,\up} = \widetilde{\chi}_{2,\down}\coloneq \widetilde{\chi}_{2}$.
The full result is shown in Fig.~\ref{fig:Hopping-elements}.
\\

\begin{figure}[H]
    \centering
    \includegraphics[width=\columnwidth]{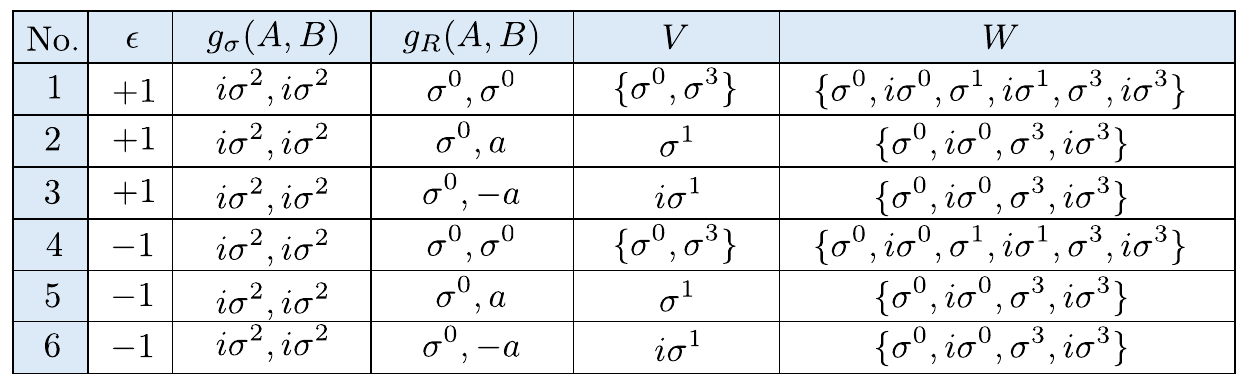}
    \caption{Modified from \cite{tarabunga2023classification}: PSG classification of parton mean-field Hamiltonians. $\epsilon=\pm 1$ denotes doubling of unit cell in parton space, $g_\sigma(A,B)$ and $g_R(A,B)$ are projective representation of reflection and rotation symmetries, $\sigma^\mu$ are Pauli matrices and $a=e^{i\frac{2\pi}{3}{\sigma^3}}$.
    $V$ and $W$ are the matrices in \eqref{eq:final-mf-hamiltonian} allowed by PSGs.
    }
    \label{fig:PSG-classification}
\end{figure}

In \cite{tarabunga2023classification}, a classification scheme of QSLs based on projective symmetry groups (PSG) \cite{Wen2002} was used to construct a parton mean-field wavefunction of the Rydberg QSL model eq. \eqref{eqn:Rydberg_BHH_fullH}.
To this end, six distinct PSGs were identified that may describe the model in the form of a parton mean-field theory. They are characterized by their SU(2) representation of reflection $g_\sigma(A,B)$ and $\pi/3$-rotation $g_R(A,B)$ and have nonvanishing mean-field amplitudes $V$ and $W$ of the Hamiltonian eq.~\ref{eq:MF-Hamiltonian-spinor-form} (see Fig.~\ref{fig:PSG-classification}).
%
%

The amplitudes $V$ and $W$ are represented as linear combinations of Pauli matrices.
For each ansatz, $V$ and $W$ were taken as variational parameters and optimized in order for the resulting ground-state to have maximum overlap with the ED wavefunction (see Sec.~\ref{sec:ED-comparison}).
The ansatz with the largest overlap with ED wavefunctions was found to be ansatz no.~1.
The optimal parameters obtained for $g=0.7$ were $V= v \cdot \sigma_0$ and $W / v =-0.31i\,\sigma_0 -0.1i\,\sigma_3$. 
The overlap obtained for these parameters is shown in Fig.~\ref{fig:Overlap-ED-MF}.
Our self-consistent solution yields ${\widetilde{\chi}_1^{(\up)}=\widetilde{\chi}_1^{(\down)}=-0.238}$ and $\widetilde{\chi}_2^{(\up)}=\widetilde{\chi}_2^{(\down)}=0.019 + i0.280$. Thus we find directly, i.e. without fitting to numerical results:
\begin{eqnarray}
    V= v \cdot\sigma_0,\qquad W/v\approx -0.26 \, (1+i)\sigma_0,
\end{eqnarray}
where $v\in \mathbb{R}$.

We see that although in general our mean-field Hamiltonian \eqref{eq:final-mf-hamiltonian} may be outside the possible ansätze since $\widetilde{\chi}_{1,\alpha}$ could in principle be complex, the self-consistently obtained mean-field amplitudes are indeed in the same ansatz class no.~1 in table \ref{fig:PSG-classification} as found in \cite{tarabunga2023classification}.

In summary, our analytical expression allows us to determine self-consistently $\widetilde{\chi}_{1/2,{\alpha}}$ and therefore the mean-field ground state. 
Our method of constructing the ground state wavefunction is not restricted 
to small system sizes in contrast to ED. (Note that in order to find the appropriate ansatz Hamiltonian in \cite{tarabunga2023classification} a comparison with ED simulations for rather small systems sizes (16 unit cells) had to be done.)

\section{Comparison to ED-wavefunctions}\label{sec:ED-comparison}

Having found the self-consistent mean-field ground-state, we can now compare it with the results obtained by ED.
Since ED can only be performed on finite lattices, we compute the mean-field ground state on finite lattices as well  \cite{silvi2013matrix}.
We then perform a Gutzwiller projection in order to obtain the corresponding bosonic wavefunction.
For $g>0$, the spectrum is gapped.
Since a chiral spin liquid is expected to be described in the low energy regime by a Chern-Simons theory \cite{bieri2016projective}, 
which implies a ground state degeneracy depending on the genus $g$ of the parameter space, we expect the ground state to show two-fold topological degeneracy.
For ED, this degeneracy could not be observed \cite{ohler2023quantum}.
To test ground-state degeneracy in the self-consistent MF approach, we compute four distinct ground states by twisting the boundary conditions $\Theta_{x,y} \in \{0, \pi\}$ for the fermions, which is explained in detail in \cite{mei2015modular}.
Two linearly independent ground states $\ket{\psi^ {(I)}}$, $\ket{\psi^ {(II)}}$ can now be obtained by computing the eigenvectors for nonvanishing eigenvalues of the overlap matrix $O_{ij}=\braket{\psi_i | \psi_j}$, where each $\ket{\psi_j}$ stands for one of the four ground states.
For $g > 0$, we observe two of these eigenvalues being zero up to a tolerance of $10^{-2}$ similar to \cite{tarabunga2023classification}.
Remarkably, this twofold topological degeneracy has not been found in ED-simulations but emerges naturally in the parton mean-field description.
\begin{figure}[H]
    \includegraphics[width=\columnwidth]{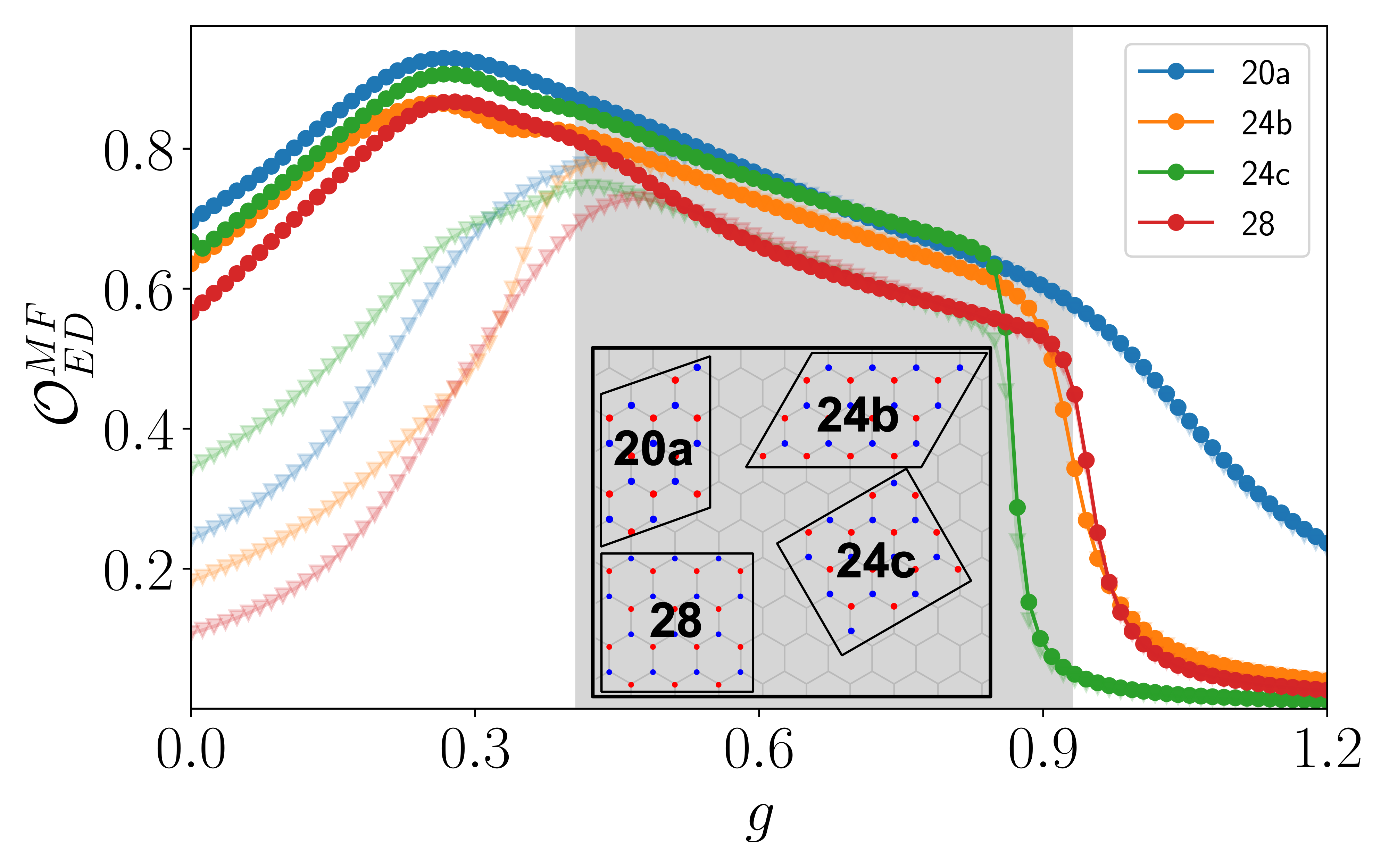}
    \caption{Overlap of parton mean-field and ED-wavefunctions for different lattice shapes. Light-colored, triangle marked values show the overlap obtained for the PSG-method as used in \cite{tarabunga2023classification}. Grey background marks the rough points of phase transition obtained from ED \cite{ohler2023quantum}. Note that the exact points of phase transition depend heavily on the shape on which ED was performed.}
    \label{fig:Overlap-ED-MF}
\end{figure}
Finally, the overlap with the ED-ground-state reads 
\begin{equation}
    \mathcal{O}^{MF}_{ED} = \sqrt{|\braket{\psi_{ED}|\psi^{(I)}_{MF}}|^2 + |\braket{\psi_{ED}|\psi^{(II)}_{MF}}|^2}.
\end{equation}
The result is shown in Fig.~\ref{fig:Overlap-ED-MF}.
The mean-field ground-state shows large overlap within the QSL-phase for ${0.4\lesssim g \lesssim 0.9}$.
Remarkably, the maximum overlap is achieved for even lower $g$-values, where ED-measurements using a ground-state fidelity metric predict the system still to be in a BEC-phase \cite{ohler2023quantum}. 
This observations could point at the existence of an additional intermediate phase already suggested in \cite{tarabunga2023classification}.
On the other hand  our mean-field model is not capable of predicting phase transitions, as we excluded spin order by setting $\langle\hat{\mathbf{S}}\rangle =0$, see eq.\eqref{eq:av-S}.
We note that for the comparably small system sizes used in the ED-simulations, a high overlap does not necessarily imply a QSL-phase, but rather that low values of $\mathcal{O}^{MF}_{ED}$ indicate the absence of a QSL-phase.   
\begin{figure}[H]
            \includegraphics[width=\columnwidth]{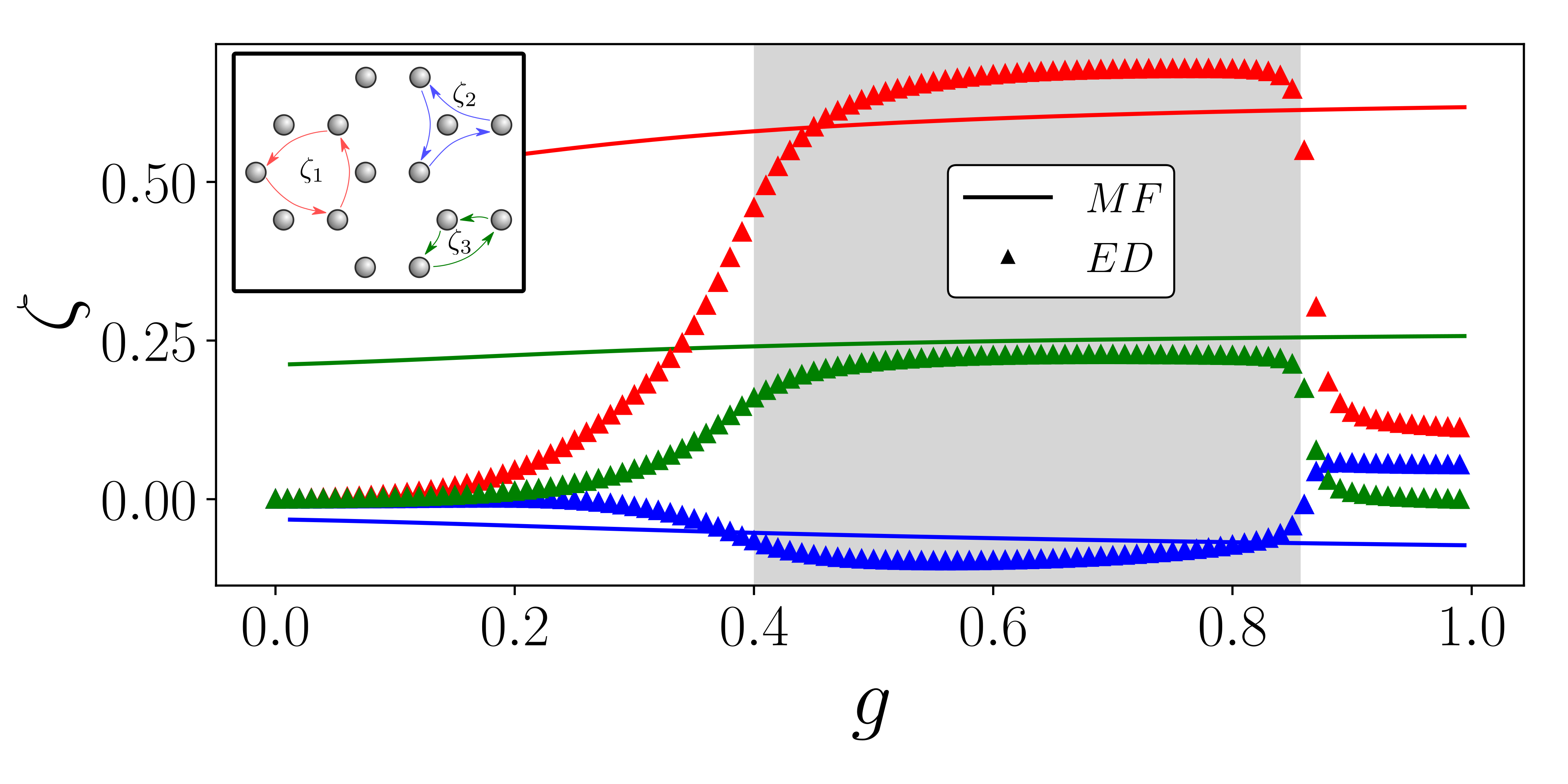}
        \caption{Spin chirality, eq. \eqref{eq:spin_chirality}, as a function of $g$ obtained within the parton MF approach (solid lines) and exact diagonalization (ED) (triangles) for the three different triangles indicated in the inset. One recognizes very good agreement within the CSL phase marked by a gray background (Note that the parton MF approach is expected to describe only the CSL phase.)
        }
        \label{fig:chirality}
\end{figure}

We also observe that the overlap of the wavefunctions for both our method and the PSG-method used in \cite{tarabunga2023classification} with the ED-wavefunctions are remarkably similar within the QSL-phase.
However, our method is now able to give an insight in the physical origin of the mean-field hopping amplitudes $V$ and $W$.

\begin{figure}[H]
    \centering
    \begin{subfigure}[t]{0.44\columnwidth}
        \includegraphics[width=\columnwidth]{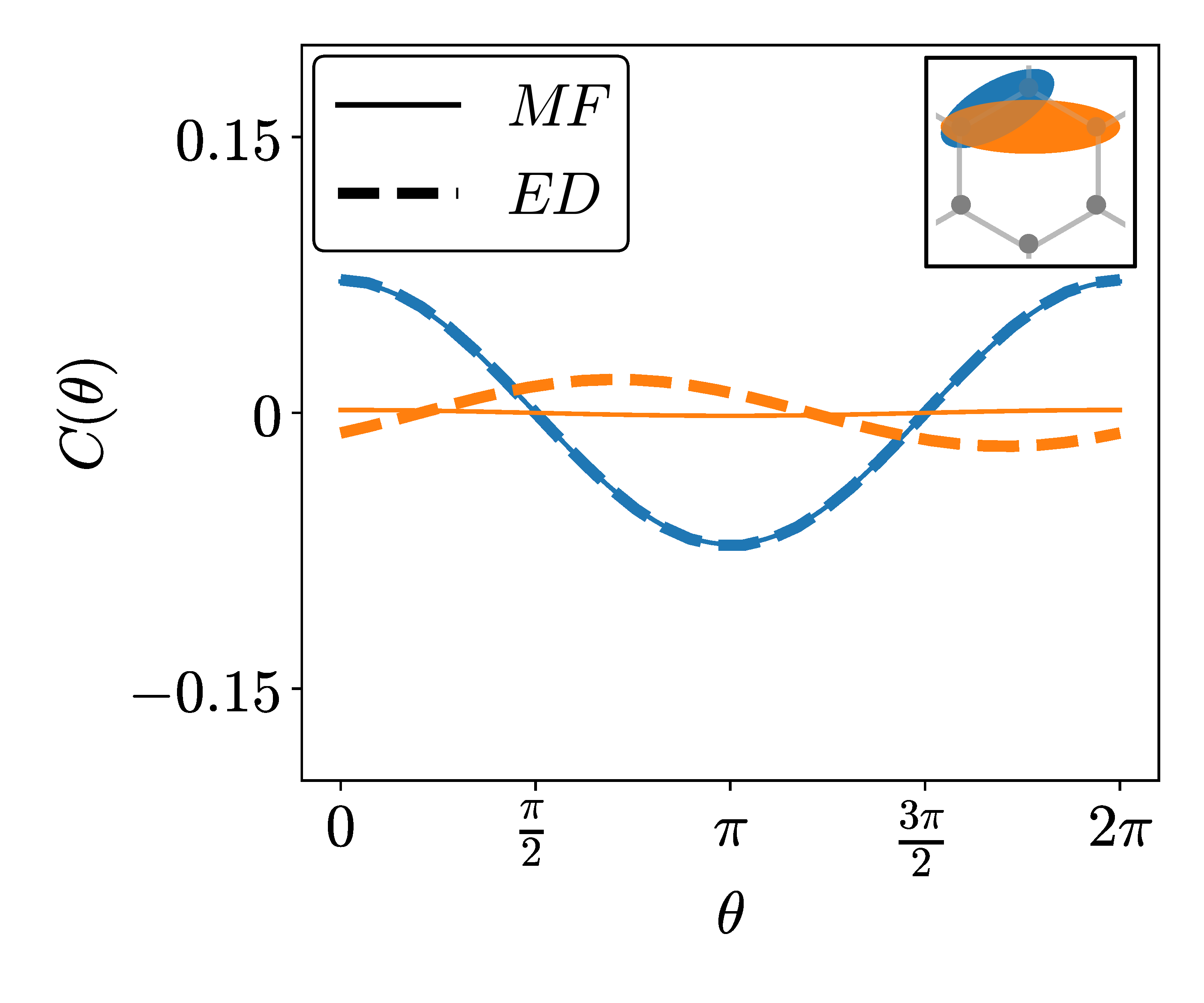}
         \caption{}
        \label{fig:spin-orientation}
    \end{subfigure}
    \quad
    \begin{subfigure}[t]{0.5\columnwidth}
        \includegraphics[width=\columnwidth]{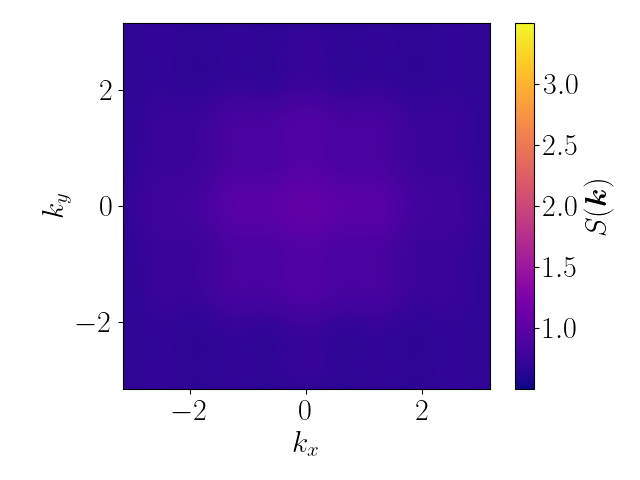}
         \caption{}
        \label{fig:spin-structure-factor}    
    \end{subfigure}
   \caption{(a) In-plane spin-correlation, eq.~\eqref{eq:spin-correlation}, from mean-field and ED simulations.
   One recognizes perfect agreement for NN (blue) and reasonable agreement for NNN (orange) spin correlations. (b) Static spin-structure factor, eq.~\eqref{eq:spin-structure} at $g=0.5$ showing absence of simple spin order.}
\end{figure}

Next we  compare the spin chirality
\begin{equation}\label{eq:spin_chirality}
    \zeta = \langle\hat{\vec{\sigma}}_i\cdot(\hat{\vec{\sigma}}_j\times\hat{\vec{\sigma}}_k)\rangle
\end{equation}
for the wave functions computed on the shape 24c, which is displayed in Fig.~\ref{fig:chirality}.
We see a very good agreement of the mean-field with the ED-result within the QSL-phase. Note again that the parton MF approach is expected to describe only the CSL phase accurately.
We can also compute other observables discussed in \cite{ohler2023quantum} that are indicators of a spin liquid phase. One such example is the in-plane spin-orientation 
\begin{equation}
C(\theta) = 4\langle \hat{S}_i^{(0)}\hat{S}_j^{(\theta)}\rangle   \label{eq:spin-correlation} 
\end{equation}
with
\begin{equation}
    \hat{S}_j^{(\theta)} = \cos(\theta)\hat{S}_i^{x} + \sin(\theta)\hat{S}_j^{y}
\end{equation}
which is shown in Fig.~\ref{fig:spin-orientation} and is identical for NNs at $g = 0.5 $.
The results for NNNs coincide well considering the already very small order of magnitude $C(\theta)\approx 0.01$.
Finally the spin-structure factor
\begin{equation}
    S(\vec{k}) = \frac{1}{N}\sum_{ij=1}^Ne^{i\vec k (\vec{r}_i - \vec{r}_j)}\langle \hat{\vec{S}}_i\cdot \hat{\vec{S}}_j\rangle \label{eq:spin-structure}
\end{equation}
for $g = 0.5$ shown in Fig.~\ref{fig:spin-structure-factor} practically vanishes, showing the same expected behavior as in the ED-simulations.
%

 \section{Conclusion}

We here discussed the ground state of Rydberg spin excitations on a honeycomb lattice with real-valued NN and complex NNN hopping processes controlled by the spin state on the intermediate lattice site. ED simulations \cite{ohler2023quantum} and subsequent symmetry considerations \cite{tarabunga2023classification} have indicated the presence of a chiral spin liquid phase induced by quantum fluctuations of the nonlinear NNN hopping at parameter values where the competition between NN and NNN terms leads to frustration. Following the fermion parton construction for spin liquids suggested in \cite{Wen2002} we here constructed a parton mean-field Hamiltonian, which is bilinear in the fermion operators but depends parametrically on fermion correlations. The ground state can then be obtained self-consistently for arbitrary system sizes. We showed that the self-consistent mean-field Hamiltonian fulfills all requirements of the classification scheme for quantum spin liquids based on projective symmetry groups. Projective symmetries alone do not completely fix the form of the parton mean-field Hamiltonian, however. 
In \cite{tarabunga2023classification} the authors postulated a specific form by maximizing the overlap of the ground state wavefunction of different mean-field Hamiltonians allowed by projective symmetries and ground states obtained by exact diagonalization of small systems. 
Our self-consistent solution yields a unique mean-field Hamiltonian which agrees with the one postulated in \cite{tarabunga2023classification} and furthermore shows large overlap with exact ground states obtained by ED on small lattices.
Finally the self-consistent solution yields a doubly degenerate ground state wavefunction as expected for chiral spin liquids, which could not be found in the ED simulations due to finite size constraints.

\subsection*{Acknowledgment}

We thank Poetri Sonya Tarabunga for very helpful and inspiring discussions. The authors gratefully acknowledge financial support from the DFG through SFB TR 185, project number 277625399. The authors also thank the Allianz f\"ur Hochleistungsrechnen (AHRP) for giving us access to the ``Elwetritsch'' HPC Cluster.

\subsection*{Author contributions}
B.B. performed the analytical calculations and the numerical mean-field simulations with support from S.O. 
S.O. implemented the ED simulations on small systems.
M.F. conceived and supervised the project.

\bibliography{Parton-Mean-Field}

\end{document}